\documentclass[%
 reprint,
 amsmath,amssymb,
 aps,
]{revtex4-1}

\usepackage{graphicx}
\usepackage{dcolumn}
\usepackage{bm}
\usepackage{amsmath,amssymb}
\usepackage{color} 

\begin{document}

\preprint{APS/123-QED}

\title{Bifurcation analysis of a density oscillator using two-dimensional hydrodynamic simulation }

\author{Nana Takeda}
\author{Naoko Kurata}
\author{Hiroaki Ito}
\author{Hiroyuki Kitahata}
\email{kitahata@chiba-u.jp}
\affiliation{
 Department of Physics, Chiba University, Chiba 263-8522, Japan
}

\date{\today}

\begin{abstract}
A density oscillator exhibits limit-cycle oscillations driven by the density difference of the two fluids.
We performed two-dimensional hydrodynamic simulations with a simple model and reproduced the oscillatory flow observed in experiments.
As the density difference is increased as a bifurcation parameter, a damped oscillation changes to a limit-cycle oscillation through a supercritical Hopf bifurcation.
We estimated the critical density difference at the bifurcation point and confirmed that the period of the oscillation remains finite even around the bifurcation point.
\end{abstract}

\maketitle


\section{introduction}

Limit-cycle oscillations often appear in various systems with energy gain and dissipation. 
Heartbeat, circadian rhythm, and firefly flashing are the famous examples in biological systems \cite{sync, Winfree, Murray}. 
Cyclic phenomena are also found in fluid systems such as geysers, thermohaline circulations, and the solar cycle \cite{flow, Wunsch}. 
These limit-cycle oscillations in nature are often complex because many factors are cooperated, and thus it is difficult to understand the underlying mechanism. 
Therefore, it should be a good approach to first understand the essential mechanism of ideal systems and then address more complex systems.

A density oscillator is an example of limit-cycle oscillators in fluid systems, first reported by Martin in 1970 \cite{Martin}.
The system consists of two different-density fluids put in two fixed containers.
The inner and outer containers are for the higher- and lower-density fluids, respectively.
The small hole at the bottom wall of the inner container connects the two fluids.
In this system, the upstream of the lower-density fluid and the downstream of the higher-density fluid alternately occur through the hole in appropriate conditions. 
Owing to the simplicity of the setup, a density oscillator has been investigated mainly in experiments \cite{Alf, Yoshikawa1, Yoshikawa2, Upa, Cerve, Ueno, Kano1, Maki, Yoshikawa3, Nakata, Miyakawa, Gonza, Horie, Steinbock, Kano2, Kano3, Ito}. 
In previous studies, it has been reported that a density oscillator shows typical characteristics of a limit-cycle oscillation such as orbital stability and synchronization among several oscillators \cite{Yoshikawa3, Nakata, Miyakawa, Gonza, Horie}.
In addition, some studies proposed theoretical models by adopting the Navier-Stokes equation for each of the upstream
and the downstream in the hole. 
Steinbock and coauthors derived the theoretical description of the critical water level in the inner containers for the reversal of the downstream in a two-dimensional system \cite{Steinbock}. 
Kano and Kinoshita focused on the intrusion of the different-density fluid in the hole and proposed a model which explained the processes of the upstream, the downstream, and the switching between them in a unified manner \cite{Kano2, Kano3}.

In spite of intensive studies on the dynamics of the flow, bifurcation structure by the change in parameters such as the hole size, the density difference between the two fluids, and the viscosity of the fluids has not been investigated in detail. To understand the dynamics of nonlinear systems, it is important to interpret the dynamics qualitatively from the viewpoint of bifurcation, where the qualitative behavior of the system changes with a bifurcation parameter. 
Our recent experiment suggested that the density oscillator shows a supercritical Hopf bifurcation between the resting and oscillatory states depending on the density difference between the two fluids \cite{Ito}. 
However, we could not definitely identify the bifurcation class only from the experimental results because the measurement of a small-amplitude oscillation around a bifurcation point suffered from relatively large error.

In order to address the identification of the bifurcation structure, numerical simulation is useful since one can accurately
quantify the dynamics in various conditions. 
In general, hydrodynamic simulation for a density oscillator has not been performed because it is difficult to treat the free surface which changes with time except for few studies: Okamura and Yoshikawa carried out a three-dimensional hydrodynamic simulation for a density oscillator with the free surface by adopting the volume of fluid method \cite{Okamura}. 
In their study, they considered essential factors in the system, e.g., the gradients of pressure, viscosity, and gravity, and proposed that the oscillation follows the Rayleigh equation. 
The simulation parameters were fixed to perform costly calculations, and the bifurcation structure remains as an open question. 
Using such ordinary differential equations as the Rayleigh equation, which were obtained by the reduction of the hydrodynamic equation, mathematical analyses were performed and the type of bifurcation was discussed \cite{Aoki, Kenfack1}. 
However, the bifurcation analysis directly based on the hydrodynamic equation has not been performed so far, which should be important to understand the actual dynamics of the density oscillator and to discuss the validity of such reductions.

In the present study, we carry out a two-dimensional hydrodynamic simulation for a density oscillator. 
In the simulation, we set the calculation area inside the fluid and associate the change in the water levels with the pressure at the calculationarea boundaries. 
Using this simple model, we obtain the time series of the density profile and the water level. 
Then we investigate the detailed bifurcation structure depending on the density difference between the two fluids.

\section{model}

We carried out a two-dimensional hydrodynamic simulation for a density oscillator.
For the incompressible viscous fluid, we use the Navier-Stokes equation and the equation of continuity as governing equations,
\begin{align}
\rho \left[ \frac{\partial \bm{v}}{\partial t} + (\bm{v} \cdot \nabla) \bm{v} \right] = - \nabla p + \mu \nabla^2 \bm{v} + \rho \bm{g},
\label{eq:Navier-Stokes equation}
\end{align}
\begin{align}
\nabla \cdot \bm{v} = 0,
\label{eq:equation of continuity}
\end{align}
where $\rho (\bm{r}, t)$ is the fluid density, $\bm{v}(\bm{r}, t) = (v_x(\bm{r}, t), v_y(\bm{r}, t))$ is the fluid velocity, $p (\bm{r}, t)$ is the pressure, $\mu$ is the fluid viscosity, and $\bm{g}$ is the acceleration of gravity.
Here, $\bm{r} = (x, y)$ is a positional vector.
For a miscible two-phase fluid, we define a normalized concentration $c(\bm{r}, t)$ $(0 \leq c \leq 1)$, where $c = 0$ and $c = 1$ correspond to the lower- and higher-density fluids, respectively.
The density is described using $c$ as
\begin{align}
\rho(\bm{r}, t) = \rho_{\rm high} + (\rho_{\rm low} - \rho_{\rm high})(1 - c(\bm{r}, t)),
\end{align}
where $\rho_{\rm low}$ and $\rho_{\rm high}$ are the densities of the lower- and higher-density fluids.
The normalized concentration $c$ satisfies an advection-diffusion equation as
\begin{align}
\frac{\partial c}{\partial t} + \nabla \cdot (c \bm{v}) = D \nabla^2 c,
\label{eq:advection-diffusion equation}
\end{align}
where $D$ is a diffusion coefficient.
We set $D$ to be small, so that the fluids hardly mix with each other \cite{Ito}.

\begin{figure}[tb]
  \begin{center}
    \includegraphics{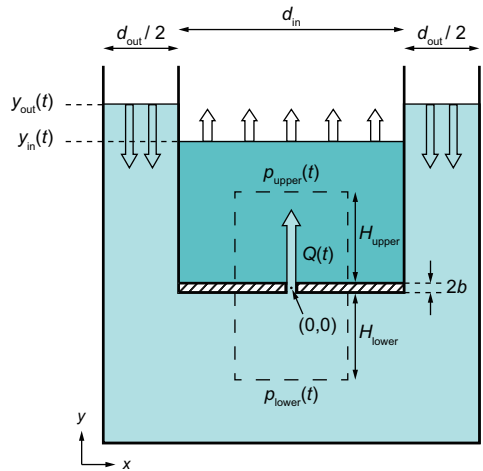}
  \end{center}
  \caption{Schematic drawing of a density oscillator when an upstream occurs. The calculation area is denoted with the broken rectangle, and the hatched area corresponds to the bottom wall of the inner container. $p_{\rm upper} (t)$ and $p_{\rm lower} (t)$ are the pressures at the upper and lower boundaries of the calculation area, respectively. $Q (t)$ is the amount of the fluid flowing through the hole per unit time. $y_{\rm in} (t)$ and $y_{\rm out} (t)$ are the water levels in the inner and outer containers, respectively.}
  \label{fig:system}
\end{figure}

Figure~\ref{fig:system} shows the schematic drawing of a density oscillator, where the calculation area is set inside the fluid.
We set the origin of the Cartesian coordinates at the center of the hole and define the area for the hole as $-a \leq x \leq a, -b \leq y \leq b$.
The calculation area is defined as $-W \leq x \leq W, -b-H_{\rm lower} \leq y \leq b+H_{\rm upper}$.
Inside it, the areas $-W \leq x < -a, -b < y < b$ and $a < x \leq W, -b < y < b$ correspond to the bottom wall of the inner container.
We assume that the pressures at the upper and lower boundaries of the calculation area, $p_{\rm upper} (t)$ and $p_{\rm lower} (t)$, are determined by the water levels in the inner and outer containers, $y_{\rm in}$ and $y_{\rm out}$, as
\begin{subequations}
	\begin{align}
p_{\rm upper} (t) = \rho_{\rm high} g \left(y_{\rm in} (t) - b - H_{\rm upper} \right),
\label{eq:pressure_waterlevel_1}
\\
p_{\rm lower} (t) = \rho_{\rm low} g \left(y_{\rm out} (t) + b + H_{\rm lower} \right),
\label{eq:pressure_waterlevel_2}
	\end{align}
\end{subequations}
where $g = |\bm{g}|$.
Here, we set the atmospheric pressure to be zero.
The change in the water level is determined by the amount of the fluid flowing through the hole per unit time $Q(t)$, which is calculated as
\begin{align}
Q(t) = \int_{-a}^{a} v_y (x, b, t) dx.
\end{align}
Due to the incompressibility, the change in the water levels can be directly obtained from $Q$ as
\begin{subequations}
	\begin{align}
\frac{dy_{\rm in}}{dt} &= \frac{Q}{d_{\rm in}},
\label{eq:inner_waterlevel}
\\
\frac{dy_{\rm out}}{dt} &= - \frac{Q}{d_{\rm out}},
\label{eq:outer_waterlevel}
	\end{align}
\end{subequations}
where $d_{\rm in}$ and $d_{\rm out}$ are the widths of the inner and outer containers, respectively, as shown in Fig.~\ref{fig:system}.

\begin{figure*}[tb]
  \begin{center}
    \includegraphics{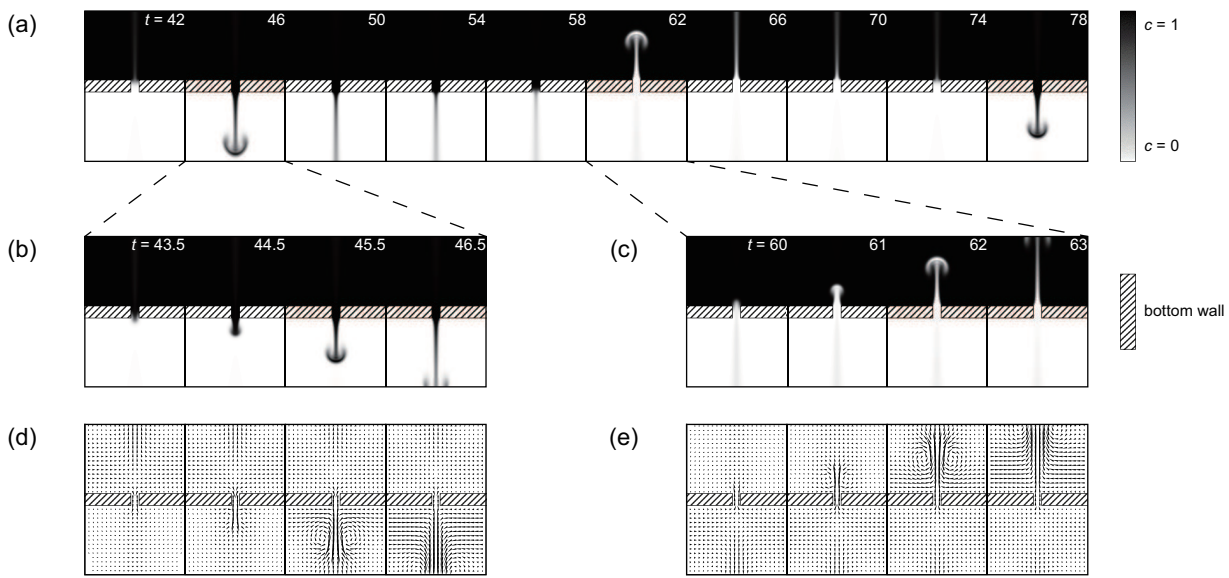}
  \end{center}
  \caption{Snapshots of density profile (a--c) and velocity field (d,e) for $\Delta \rho = 0.2$. (a) Typical time series of the oscillatory flow. (b,d) Detailed time series of downstream. (c,e) Detailed time series of upstream. The hatched area corresponds to the bottom wall of the inner container.}
  \label{fig:profile}
\end{figure*}

In this way, we associate the change in the water levels with the pressure at the calculation-area boundaries, and obtain a simple model, where the free surface need not be directly dealt with.
A non-slip boundary condition for the velocity, $\bm{v} = \bm{0}$, and the Neumann boundary condition for the density, $\nabla_\perp \rho = 0$, are set for the surfaces of the bottom wall, where $\nabla_\perp$ represents the derivative in the normal direction.
The pressure at the surfaces of the bottom wall is determined so that it satisfies the Navier-Stokes equation in Eq.~(\ref{eq:Navier-Stokes equation}).
The pressure at the upper and lower boundaries of the calculation area follows Eqs.~(\ref{eq:pressure_waterlevel_1}) and (\ref{eq:pressure_waterlevel_2}).
The velocity, the pressure, and the density at the other calculation-area boundaries also follow the Neumann condition.
At the initial state, the two fluids are stationary ($\bm{v} = \bm{0}$) and not mixed, where the concentration is set as $c = 0$ ($y < b$) and $c = 1$ ($y \geq b$).
The initial pressure of the lower- and higher-density fluids, $p_{\rm low, 0} (x, y)$ and $p_{\rm high, 0} (x, y)$, are given as
\begin{subequations}
	\begin{align}
p_{\rm low, 0} (x, y) &= \rho_{\rm low} g \left(y_{\rm out} (0) - y \right),
\\
p_{\rm high, 0} (x, y) &= \rho_{\rm high} g \left(y_{\rm in} (0) - y \right),
	\end{align}
\end{subequations}
where $y_{\rm out} (0)$ and $y_{\rm in} (0)$ are the initial water levels in the outer and inner containers, respectively.
$y_{\rm in} (0)$ is determined by the gravitational equilibrium as $y_{\rm in}(0) = b + (\rho_{\rm low} / \rho_{\rm high}) (y_{\rm out}(0) - b)$.
In the simulation, we represent the surface height of the fluid in the outer container defined as
\begin{align}
h(t) = y_{\rm out} (t) - y_{\rm out} (0),
\end{align}
where the initial water level in the outer container $y_{\rm out} (0)$ is not relevant to the behavior of the fluid because only the time derivatives of the water levels are included in the governing equations in Eqs.~(\ref{eq:inner_waterlevel}) and (\ref{eq:outer_waterlevel}).

To solve the Navier-Stokes equation (Eq.~(\ref{eq:Navier-Stokes equation})) and the equation of continuity (Eq.~(\ref{eq:equation of continuity})), we used the Marker and Cell method and calculated the velocity and the pressure on a staggered grid \cite{Mac1,Mac2}.
We used an explicit method for the advection-diffusion equation (Eq.~(\ref{eq:advection-diffusion equation})) to calculate the time evolution of $c$.
Here, we set the time step $dt = 0.0002$, the spatial mesh $dx = dy = 0.005$, and the parameters for the calculation area $W = 0.4$, $H_{\rm upper} = H_{\rm lower} = 0.55$.
The simulation parameters were set as follows: $\mu = 1/300, D = 0.0001, g = 10, a = 0.03, b = 0.05$, $d_{\rm in} = d_{\rm out} = 4.8$, and $y_{\rm out}(0) = 10.05$.
To investigate the bifurcation structure, we changed the density difference $\Delta \rho \, (= \rho_{\rm high} - \rho_{\rm low})$ as a bifurcation parameter, where we fixed $\rho_{\rm low} = 1$ and varied $\rho_{\rm high}$.

\section{simulation results}

The snapshots of the density profile and the velocity field for the density difference $\Delta \rho = 0.2$ are shown in Fig.~\ref{fig:profile}.
In one period, the downstream occurs ($t = 43.5$) and passes through the lower boundary of the calculation area ($t = 46.5$).
After a while, the downstream becomes weaker and changes to the upstream ($t = 60$).
Then the upstream grows and passes through the upper boundary of the calculation area ($t = 63$).
The upstream becomes weaker and changes to the downstream again.
In this way, the limit-cycle oscillation consisting of the upstream and the downstream occurs with a period of $\sim 30$.

\begin{figure}[tb]
  \begin{center}
    \includegraphics{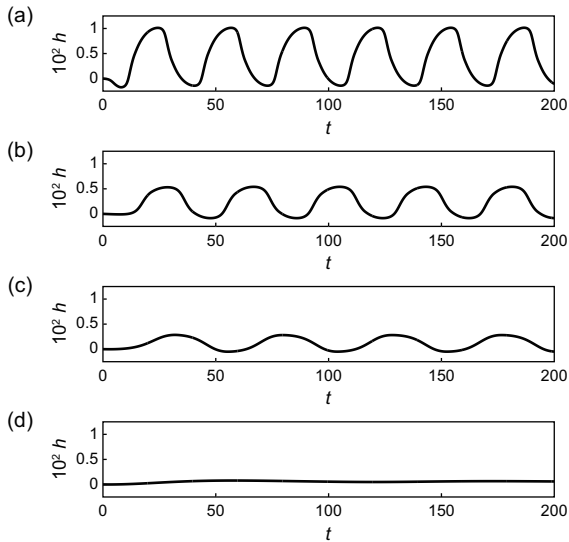}
  \end{center}
  \caption{Time series of $h$ for (a) $\Delta \rho = 0.2$, (b) $\Delta \rho = 0.1$, (c) $\Delta \rho = 0.05$, and (d) $\Delta \rho = 0.025$.}
  \label{fig:wave}
\end{figure}

\begin{figure}[tb]
  \begin{center}
    \includegraphics{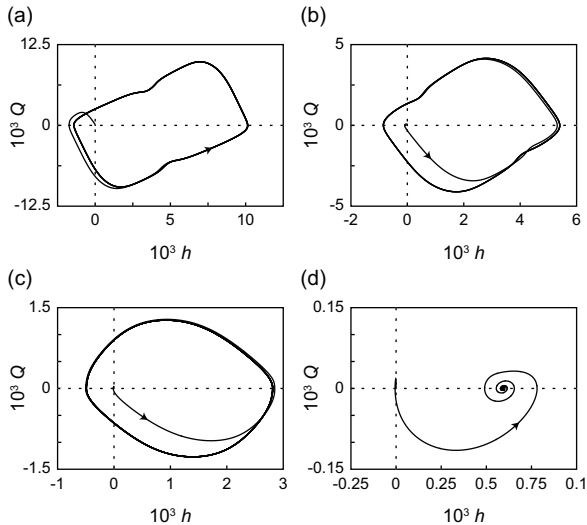}
  \end{center}
  \caption{Trajectories in a phase space of $h$ and $Q$ for (a) $\Delta \rho = 0.2$ $(0 \leq t \leq 200)$, (b) $\Delta \rho = 0.1$ $(0 \leq t \leq 200)$, (c) $\Delta \rho = 0.05$ $(0 \leq t \leq 600)$, and (d) $\Delta \rho = 0.025$ $(0 \leq t \leq 600)$.}
  \label{fig:phase space}
\end{figure}

Figure~\ref{fig:wave} shows the changes in the surface height in the outer container $h$.
Figure~\ref{fig:wave}(a) shows a relaxation oscillation in which $h$ asymptotically approaches the two heights alternately.
As the density difference decreases, the waveform around the peaks changes as shown in Fig.~\ref{fig:wave}(b).
The system does not exhibit a relaxation oscillation but exhibits a harmonic-like oscillation as shown in Fig.~\ref{fig:wave}(c).
In addition to the waveform change, the amplitude of the oscillation decreases, and the period increases.
For the smaller density difference, the oscillation gets damped as shown in Fig.~\ref{fig:wave}(d).
Figure~\ref{fig:phase space} shows the trajectories in a phase space of $h$ and the flow rate through the hole $Q$, which corresponds to $- d_{\rm out} dh / dt$.
For $\Delta \rho = 0.2, 0.1,$ and $0.05$, the trajectory converges to a closed orbit, suggesting that the system exhibits a limit-cycle oscillation as shown in Figs.~\ref{fig:phase space}(a--c).
For $\Delta \rho = 0.025$, the trajectory converges to a fixed point, and the system shows a damped oscillation as shown in Fig.~\ref{fig:phase space}(d).
These results reveal that a density oscillator shows bifurcation from the resting to oscillatory states depending on the density difference $\Delta \rho$.

\begin{figure}[tb]
  \begin{center}
    \includegraphics{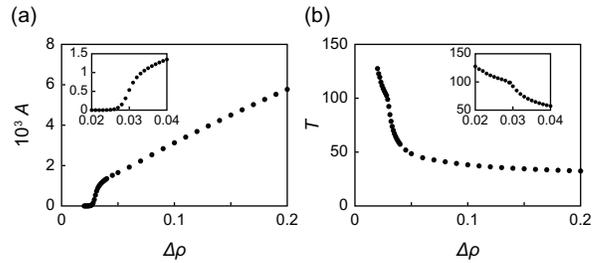}
  \end{center}
  \caption{Bifurcation diagram of the density oscillator. (a) Amplitude $A$ and (b) period $T$ of the oscillation of $h$ depending on $\Delta \rho$. Each inset shows the expanded area from $\Delta \rho = 0.02$ to $\Delta \rho = 0.04$.}
  \label{fig:bifurcation}
\end{figure}

To obtain the amplitude and the period of the oscillation of $h$ for each $\Delta \rho$, we detected the instances with $Q=0$ and set the $i$-th instance as $t_i$ ($i = 1, 2, \cdots$).
$t_0$ was set to be $0$ since $Q = 0$ in the initial condition.
The $n$-th period was defined as $T_n = t_{2n} - t_{2n-2}$ and the $n$-th amplitude was defined as $A_n = \left[|h(t_{2n-1}) - h(t_{2n-2})| + |h(t_{2n-1}) - h(t_{2n})|\right]/4$ ($n = 1, 2, \cdots$).
Then we defined the amplitude $A$ and the period $T$ as $A = (A_{\rm 4} + A_{\rm 5}) / 2$ and $T = (T_{\rm 4} + T_{\rm 5}) / 2$, since $A_{n}$ and $T_{n}$ for $n = 1, 2,$ and $3$ were not used to avoid the initial value dependence.
Figure~\ref{fig:bifurcation} shows the amplitude $A$ and the period $T$ of the oscillation of $h$ for $0.02 \leq \Delta \rho \leq 0.2$.
It should be noted that the system for $\Delta \rho = 0.025$ exhibits a damped oscillation as shown in Figs.~\ref{fig:wave}(d) and \ref{fig:phase space}(d), and thus the region of $\Delta \rho$ plotted in Fig.~\ref{fig:bifurcation} reflects the behaviors for $\Delta \rho$ in the both sides of the bifurcation point.
With a decrease in the density difference, the amplitude steeply decreases to zero for $\Delta \rho \lesssim 0.03$, which suggests the existence of a bifurcation point $\Delta \rho_{\rm c}$ as shown in Fig.~\ref{fig:bifurcation}(a).
The period is large especially around the bifurcation point, while it clearly remains finite and does not diverge as shown in Fig.~\ref{fig:bifurcation}(b).
Since the resting state changes to the oscillatory state with a finite period according to the increase in the density difference as a bifurcation parameter, the bifurcation is classified into the supercritical Hopf bifurcation.

\section{discussion}

\begin{figure}[tb]
  \begin{center}
    \includegraphics{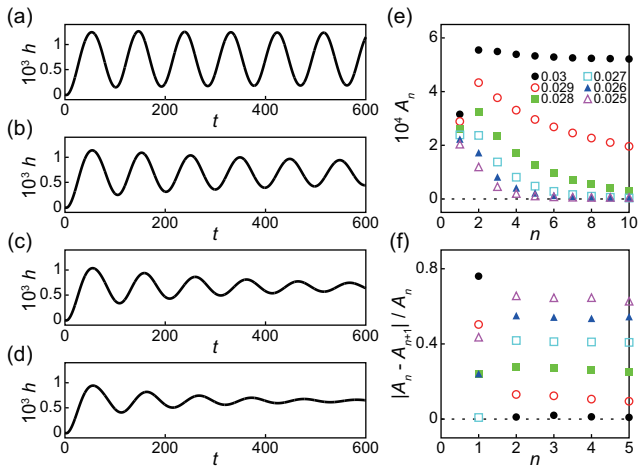}
  \end{center}
  \caption{Time series of $h$ for (a) $\Delta \rho = 0.03$, (b) $\Delta \rho = 0.029$, (c) $\Delta \rho = 0.028$, and (d) $\Delta \rho = 0.027$. (e) Change in the amplitude $A_{n}$ plotted against $n$ for $\Delta \rho = 0.03$ ($\bullet$; black), 0.029 ($\circ$; red), 0.028 ($\blacksquare$; green), 0.027 ($\square$; cyan), 0.026 ($\blacktriangle$; blue), and 0.025 ($\vartriangle$; magenta). (f) Damping rate $|A_{n} - A_{n + 1}| / A_{n}$ plotted against $n$ for each $\Delta \rho$. The symbols correspond to those in (e).}
  \label{fig:damped}
\end{figure}

The simulation results show that a density oscillator exhibits the supercritical Hopf bifurcation with the change in the density difference as a bifurcation parameter.
In Fig.~\ref{fig:bifurcation}(a), we evaluated the amplitude from the finite-time behaviors.
In general, the convergence to a fixed point or a limit cycle takes long time around the bifurcation point.
Thus, the amplitude $A$ calculated only from $A_4$ and $A_5$ could have a larger value than the amplitude of a limit cycle for $\Delta \rho \geq \Delta \rho_{\rm c}$ and $0$ for $\Delta \rho \leq \Delta \rho_{\rm c}$.
Here, we investigate the behavior of the amplitude and the period around the bifurcation point in detail, and we estimate the critical density difference at the bifurcation point $\Delta \rho_{\rm c}$.
Figures~\ref{fig:damped}(a--d) show the changes in the surface height in the outer container $h$ for $\Delta \rho = 0.027, 0.028, 0029,$ and $0.03$ around the bifurcation point.
Due to the slow convergence, we could not determine whether the system shows a limit-cycle oscillation or a damped oscillation only from the apparent waveforms within a finite duration.
To judge the types of the oscillation, we check the time evolution of the $n$-th amplitude $A_{n}$ for each $\Delta \rho$ as shown in Fig.~\ref{fig:damped}(e).
$A_{n}$ should converge to a nonzero value for a limit-cycle oscillation and to $0$ for a damped oscillation as a long-time behavior.
We also evaluate a damping rate $|A_{n} - A_{n + 1}| / A_{n}$ as shown in Fig.~\ref{fig:damped}(f).
It should converge to $0$ for a limit-cycle oscillation and to a nonzero value for a damped oscillation under the following assumption; if we assume the amplitude of a linear damped oscillation represented by $A(t) = A(0)e^{- \gamma t}$, a damping rate is calculated as $|A_{n} - A_{n + 1}| / A_{n} = 1 - e^{- \gamma T}$, where $\gamma$ is a damping coefficient.
A damping rate converges to a nonzero value between 0 and 1 because the period $T$ has a constant value.
The results shown in Figs.~\ref{fig:damped}(e,f) suggest that the system exhibits a limit-cycle oscillation for $\Delta \rho \geq 0.03$ and a damped oscillation for $\Delta \rho \leq 0.027$, and there is a bifurcation point between them.

\begin{figure}[tb]
  \begin{center}
    \includegraphics{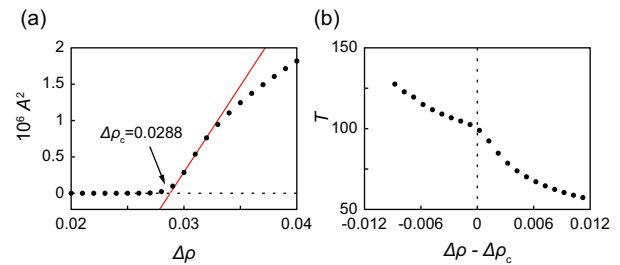}
  \end{center}
  \caption{Critical density difference at the bifurcation point $\Delta \rho_{\rm c}$ and the period around $\Delta \rho_{\rm c}$. (a) Squared amplitude $A^2$ depending on $\Delta \rho$ and linear fitting with the three values $\Delta \rho = 0.03, 0.031,$ and $0.032$. (b) Period $T$ depending on $\Delta \rho - \Delta \rho_{\rm c}$.}
  \label{fig:fitting}
\end{figure}

It is known that the amplitude of the oscillation for the supercritical Hopf bifurcation follows the scaling of $(\Delta \rho - \Delta \rho_{\rm c})^{1/2}$ \cite{Strogatz}.
Figure~\ref{fig:fitting}(a) shows the squared amplitude $A^2$ plotted against $\Delta \rho$ and the linear fitting by a least squares method with the three values $\Delta \rho = 0.03, 0.031,$ and $0.032$, which belong to $\Delta \rho$ for a limit-cycle oscillation close to the bifurcation point.
For $\Delta \rho \geq 0.033$, the values of $A^2$ gradually deviate from the linear fitting.
The region of the linear fitting close to the bifurcation point means that the amplitude $A$ increases from 0 with the scaling of $(\Delta \rho - \Delta \rho_{\rm c})^{1/2}$.
We estimate $\Delta \rho_{\rm c} = 0.0288$ from the intersection of the linear fitting and $A^2 = 0$.
Figure~\ref{fig:fitting}(b) shows the period $T$ plotted against $\Delta \rho - \Delta \rho_{\rm c}$, where $\Delta \rho_{\rm c}$ is estimated in Fig.~\ref{fig:fitting}(a).
The slope of the period changes at around the bifurcation point $\Delta \rho = \Delta \rho_{\rm c}$, and the period has finite values for $\Delta \rho \leq \Delta \rho_{\rm c}$ as well as $\Delta \rho \geq \Delta \rho_{\rm c}$, which corresponds to the behavior of the supercritical Hopf bifurcation.

The simulation result agrees well with our previous experimental result, where pure water and a sodium chloride aqueous solution were adopted as the lower and higher density solutions, respectively \cite{Ito}. 
Actually, we obtained the bifurcation diagram in Fig.~\ref{fig:bifurcation} qualitatively consistent with that obtained from the experiment, where the amplitude increased, and the period decreased, according to the increase in the density difference above the bifurcation point. 
The amplitude increased from the bifurcation point approximately with the scaling of $(\Delta \rho - \Delta \rho_{\rm c})^{1/2}$ both in the simulation and in the experiment. 
Leaving the region close to the bifurcation point, these systems exhibited slightly different behaviors. 
In Fig.~\ref{fig:fitting}(a) the values of the squared amplitude $A^2$ gradually deviated downward from the linear fitting with an increase in the density difference, whereas they deviated upward in the experimental result. 
Despite the difference in the behaviors far from the bifurcation point, it is notable that the simulation and experimental results exhibited the same bifurcation structure in the region around the bifurcation point.

Here we discuss the validity of the two-dimensional simulation compared with the three-dimensional experimental system.
It is expected that the fluid amount flowing through the hole and the resistance inside the hole in the two-dimensional system are quantitatively different from those in the three-dimensional system.
In spite of these quantitative differences, bifurcation structures in the both systems can be discussed in the same framework as follows.
The existence of a critical density difference is interpreted from the competition between the diffusion and advection of the solute.
To evaluate the critical density difference $\Delta \rho_c$ in the experiment, we considered the Rayleigh number $Ra = (gL^3\Delta\rho/\rho)/(\nu D)$, where $L$ is a characteristic length, and $\nu = \mu / \rho$ is the kinematic viscosity of the fluid.
We adopted $Ra \sim 10^3$, which is typical for Rayleigh-B\'{e}nard instability, and the estimated value of $\Delta \rho_c$ agreed well to the experimental result \cite{Ito}.
The critical density difference is estimated also by the simulation parameters: $g = 10$, $L \sim 0.1$, $\mu = 1/300$, and $D = 0.0001$.
Here, we use the length scale of the hole size as $L$.
The estimated value $\Delta \rho_c \sim 0.03$ approximately agrees with the order of the estimation by the simulation result in Fig.~\ref{fig:fitting}(a).
The above discussion from the viewpoint of the transport phenomenon suggests that the bifurcations between the resting and oscillatory states originate from the same mechanism, regardless of the difference in dimensionality.

In previous studies, the ordinary differential equations reduced from a hydrodynamic equation were often adopted as simple mathematical models for the density oscillators, and the detailed bifurcation structures for the mathematical model have been investigated. 
For example, Aoki proposed a model with a nonlinear frictional term and claimed that the system has three fixed points. He investigated the stability of each point depending on the bifurcation parameter, which accelerates the fluid velocity around the orifice, and discussed the bifurcation structure \cite{Aoki}. 
Kenfack et al. also derived the piecewise-smooth ordinary differential equations and claimed that the system exhibits the nonconventional Hopf bifurcation by varying the resistance coefficient, determined by the viscosity and the geometry of the orifice, as a bifurcation parameter \cite{Kenfack1, Kenfack2}. 
It is difficult to directly discuss the correspondence between their models and our simulation result because their bifurcation parameters are essentially different from ours, i.e., the density difference. 
We guess that their results on the detailed bifurcation structure may come from the assumption or approximation in the reduction processes.

In  an actual density oscillator, the higher- and lower-density fluids mix, and the amplitude and the equilibrium height changes with time.
In our model, we assume that the density in the outside of the calculation area is constant. 
Strictly, it should not be constant because the different-density fluid goes through the calculation boundaries. 
Considering that only the derivative of $y_{\rm out}(t)$ is important and the absolute value of $y_{\rm out}(t)$ does not matter owing to Eqs.~(\ref{eq:inner_waterlevel}) and (\ref{eq:outer_waterlevel}), the size outside of the calculation area can be set arbitrarily.
Therefore, our model can be regarded as an ideal system in which the mixing effect between the higher- and lower-density fluids is negligible independently of the choice of $y_{\rm out}(0)$.

\section{conclusion}

We carried out a two-dimensional hydrodynamic simulation for a density oscillator using a simple model associating the change in the water levels with the pressure at the calculation-area boundaries.
Using this simulation, we clarified that a density oscillator shows supercritical Hopf bifurcation between a damped oscillation and a limit-cycle oscillation depending on the density difference $\Delta \rho$.
We confirmed that the amplitude increases from 0 with the scaling of $(\Delta \rho - \Delta \rho_{\rm c})^{1/2}$, and the period has a finite value at the bifurcation point estimated from the scaling of the amplitude.
The simulation of a density oscillator can be useful for the investigation of other phenomenon in density oscillators such as synchronizations and phase responses.
The present study will contribute to further understanding of the nonlinear phenomenon in oscillators in fluidic systems, as well as a density oscillator.

\begin{acknowledgments}
This work was supported by JSPS KAKENHI Grants No. JP19H00749, No. JP19K14675, and No. JP16H03949. 
It was also supported by Sumitomo Foundation (No. 181161), the Japan-Poland Research Cooperative Program “Spatiotemporal patterns of elements driven by self-generated, geometrically constrained flows,” and the Cooperative Research Program of “Network Joint Research Center forMaterials and Devices” (No. 20194006).
\end{acknowledgments}

\end{document}